\documentclass[final,3p,12pt,times]{elsarticle}

\usepackage{amssymb}
\usepackage{amsmath}
\usepackage{amsthm}
\usepackage{xcolor}%

\usepackage{lineno}
\usepackage{url}
\usepackage{hyperref}

\journal{Chaos, Solitons and Fractals}

\begin{document}

\begin{frontmatter}



\title{Predicting the onset of period-doubling bifurcations via dominant eigenvalue extracted from autocorrelation}


%

%

\author[1,label1]{Zhiqin Ma}
\fntext[label1]{ORCID: https://orcid.org/0000-0002-5809-464X (Zhiqin Ma)}

\author[1]{Chunhua Zeng\corref{cor1}}
\ead{zchh2009@126.com}
\cortext[cor1]{Corresponding author}

\author[2,3,4]{Ting Gao}

\author[5,6]{Jinqiao Duan}

\affiliation[1]{organization={Faculty of Science},
            addressline={Kunming University of Science and Technology},
            city={Kunming},
            postcode={650500},
            country={China}}

\affiliation[2]{organization={School of Mathematics and Statistics},
addressline={Huazhong University of Science and Technology},
city={Wuhan},
postcode={430074},
country={China}}
\affiliation[3]{organization={Center for Mathematical Science},
addressline={Huazhong University of Science and Technology},
city={Wuhan},
postcode={430074},
country={China}}
\affiliation[4]{organization={Steklov‐Wuhan Institute for Mathematical Exploration},
addressline={Huazhong University of Science and Technology},
city={Wuhan},
postcode={430074},
country={China}}

\affiliation[5]{organization={Guangdong Provincial Key Laboratory of Mathematical and Neural Dynamical Systems},
addressline={Great Bay University},
city={Dongguan},
postcode={523000},
country={China}}
\affiliation[6]{organization={Department of Mathematics and Department of Physics},
addressline={Great Bay University},
city={Dongguan},
postcode={523000},
country={China}}

\begin{abstract}

Predicting the occurrence of transitions in the qualitative dynamics of many natural systems is crucial, yet it remains a challenging task. Generic early warning signals like variance and lag-1 autocorrelation identify critical slowing down near tipping points but lack practical thresholds for predicting imminent transitions. More recent studies found that the dynamical eigenvalue is rooted in the framework of empirical dynamical modeling and then estimates the dominant eigenvalue of a system from time series, providing a threshold ($|$DEV$|$ = 1) to predict bifurcations and classify their types. However, its application requires careful calibration of the hyperparameters and focuses on reconstructing system dynamics directly from data. Here, we employ Ornstein–Uhlenbeck process to derive analytic approximations for the lag-$\tau$ autocorrelation function prior to period-doubling bifurcation thereby estimating the dominant eigenvalue of dynamical systems, named dominant eigenvalue extracted from autocorrelation (DE-AC), and revealing its dynamic behaviour when approaching a period-doubling bifurcation. Theoretically, dominant eigenvalue tends to $-1$ when the system approaches a period-doubling bifurcation. In particular, we evaluated DE-AC on simulation data from cardiac alternans model and on experimental data from chick heart aggregates undergoing a period-doubling bifurcation. DE-AC reliably detected the beginning of the cardiac arrhythmia (period-doubling bifurcation) in most cases. Moreover, it demonstrated superior sensitivity and specificity as an early warning signal compared to the three widely used indicators---variance, lag-1 autocorrelation, and dynamical eigenvalue. Our theoretical and empirical results suggest that DE-AC represents a quantitative measure for predicting the onset of potentially dangerous alternating rhythms in the heart. The ability to better infer, detect, and distinguish the nature of impending transitions in complex systems will help humans manage critical transitions in biological systems.

\end{abstract}


\begin{highlights}
\item We have developed a quantitative measure, called the dominant eigenvalue extracted from autocorrelation (DE-AC), to detect how close a system is to an impending transition.
\item We employ Ornstein–Uhlenbeck process to derive analytic approximations for the lag-$\tau$ autocorrelation function prior to period-doubling bifurcation, thereby estimating the dominant eigenvalue of dynamical systems and revealing its dynamic behaviour when approaching a period-doubling bifurcation.
\item We tested DE-AC approach on both simulated and experimental data, and evaluated its performance relative to dynamical eigenvalue, lag-1 autocorrelation, and variance. DE-AC provides a reliable indicator for detecting the onset of the cardiac arrhythmia (a period-doubling bifurcation). Moreover, it demonstrated superior sensitivity and specificity as an early warning signal compared to the three widely used indicators---dynamical eigenvalue, lag-1 autocorrelation, and variance.

\end{highlights}

\begin{keyword}

period-doubling bifurcation, cardiac arrhythmia, early warning signal, dominant eigenvalue, critical slowing down




\end{keyword}

\end{frontmatter}


\section{Introduction}

Many real-world systems exhibit critical thresholds, known as tipping points, at which systems abruptly shift to the qualitatively different states~\cite{scheffer2020critical,van2016you,bi2024folding,panahi2023rate,murphy2024information}. In biology, such transitions are linked to events like cardiac arrhythmia~\cite{bury2023predicting,glass2022clocks}, depression~\cite{beck2009depression,van2014critical}, epileptic seizures~\cite{jirsa2014nature,mcsharry2003prediction,kramer2012human}, alzheimer’s diseas~\cite{zhang2025action,scheltens2021alzheimer}, asthma attacks~\cite{sandberg2000role,venegas2005self}, and microbiome dysregulation~\cite{lahti2014tipping,li2023role}. The occurrence of alternating cardiac rhythms marks the transition from normal heartbeat to arrhythmic patterns~\cite{pastore1999mechanism,gizzi2013effects}. Cardiac arrhythmia can drive the heart into a state characterized by accelerated rhythms, alternating rhythms, bursting rhythms, or even chaotic dynamics, significantly raising the risk of sudden cardiac death unless addressed promptly~\cite{quail2015predicting,watanabe2001mechanisms,sato2013formation,rosenbaum1994electrical,verrier2011microvolt}. These events are known as critical transitions, which are characterized by an abrupt shift to a new dynamical state. Therefore, it is important to anticipate the onset of critical transitions in biological systems.



\begin{figure*}[tbh]
\centering
\includegraphics[width=1.0\textwidth]{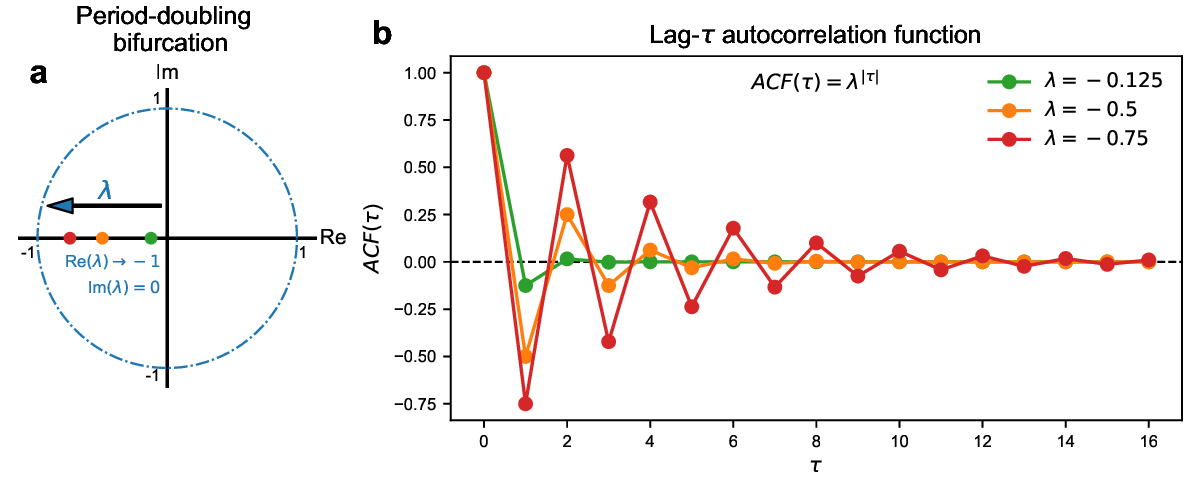}
\caption{\textbf{Analytical approximation for early warning signals preceding the period-doubling bifurcation.} (\textbf{a}) Schematic illustration of the eigenvalues in complex plane at different proximities points for the period-doubling bifurcation. The bifurcation is mathematically identified by changes in the dominant eigenvalue, $\lambda$, of the Jacobian matrix, $J$, at the tipping point. (\textbf{b}) Analytical approximation of the lag-$\tau$ autocorrelation function, $ACF(\tau)$, as the system approaches a period-doubling bifurcation. Analytical approximations for the $ACF(\tau)$ are drawn at different distances to period-doubling bifurcation given by $\lambda = \{-0.125, -0.5, -0.75\}$, where $\lambda$ is the real part of dominant eigenvalue of the system.}\label{fig1}
\end{figure*}

Cardiac rhythm transitions are marked by the emergence of alternating features within the temporal dynamics of cardiac activity, often serving as indicators for the onset of serious arrhythmic events such as ventricular tachycardia, fibrillation, and T-wave alternans~\cite{echebarria2002instability,echebarria2007amplitude}. These transitions are more clearly interpreted through bifurcation theory, a mathematical framework that investigates how dynamical systems may experience abrupt qualitative shifts when specific parameters surpass critical thresholds (known as bifurcations)~\cite{bury2023predicting,kuznetsov1998elements,strogatz2001nonlinear}. Such transitions can be linked to a mathematical instability termed a period-doubling bifurcation, which takes place when the real part of the dominant eigenvalue of a dynamical system tends to $-1$ (see Fig.~\ref{fig1}a), signaling a transition from stable periodic behavior to alternans~\cite{grziwotz2023anticipating,bury2020detecting}. This process is typically marked by critical slowing down, where the system’s local stability diminishes, resulting in systematic alterations in the properties of a noisy time series, including increased variance and autocorrelation~\cite{bury2023predicting,wissel1984universal,wiesenfeld1985noisy,scheffer2009early}. While early warning signals (EWS) like variance and lag-1 autocorrelation are proposed for identifying critical slowing down near bifurcations, their utility is constrained by qualitative trends and the absence of robust thresholds to predict imminent transitions.



Recent work integrating dynamical systems theory with machine learning offers promising avenues for EWS refinement~\cite{bury2023predicting,deb2022machine,bury2021deep,ma2025predicting,zhang2025early}. Deep learning classifiers trained on bifurcation libraries can distinguish fold, Hopf, and period-doubling transitions by detecting higher-order dynamical features. However, these approaches typically rely on system-specific training data, which is often unavailable in critical transition scenarios. More recent studies found that the dynamical eigenvalue is rooted in the framework of empirical dynamical modeling and then estimates the dominant eigenvalue of a system from time series, providing a threshold ($|$DEV$|$ = 1) to predict bifurcations and classify their types~\cite{grziwotz2023anticipating}. Specifically, the dynamical eigenvalue method uses state space reconstruction (SSR) and S-map algorithms to infer local Jacobian matrices from time series, and computes dominant eigenvalue with this information. However, its application requires careful calibration of the hyperparameters, including the embedding dimension, the time lag, and the S-map nonlinearity parameter. Crucially, dynamical eigenvalue focuses on reconstructing system dynamics directly from data rather than assuming pre-defined models, which results in a lack of mechanism description.




Here, we derive the lag-$\tau$ autocorrelation function near period-doubling bifurcation by employing Ornstein–Uhlenbeck process, analytically linking eigenvalue behavior to pre-bifurcation dynamics. Specifically, the eigenvalue was derived from the analytical expression of the lag-$\tau$ autocorrelation function, revealing its dynamic behavior as it approaches a period-doubling bifurcation. The derivations rely on standard stochastic process theory~\cite{bury2020detecting,gardiner1985handbook}, assuming quasi-stationarity (where the bifurcation parameter changes sufficiently slowly) and small noise (which makes nonlinear terms negligible). Over time, the dominant eigenvalue ($\lambda$) shifts from the interior to the border of the unit circle, signaling a critical transition, namely the onset of the period-doubling bifurcation (see Fig.~\ref{fig1}a). In Fig.~\ref{fig1}b, damped oscillations in the lag-$\tau$ autocorrelation function ($ACF(\tau)$) are we observed as a function of lag time ($\tau$). As $\lambda$ tends to $-1$, the autocorrelation function decays more slowly to zero (indicating no correlation), and the amplitude of the oscillations increases. When the system is far from the bifurcation (e.g., $\lambda = -0.125$, green), the autocorrelation function is uncorrelated, thus the lag-$\tau$ autocorrelation function is nearly zero. As the system approaches the period-doubling bifurcation, the lag-$\tau$ autocorrelation function exhibits damped oscillations, which reflect the prolonged recovery times after perturbations near the period-doubling bifurcation, as shown in the ($\lambda=-0.5$, orange) and ($\lambda=-0.75$, red) of Fig.~\ref{fig1}b. In summary, in the vicinity  of a period-doubling bifurcation, critical slowing down occurs, causing the system takes longer to recover to equilibrium after a perturbation.

To validate the effectiveness of dominant eigenvalue extracted from autocorrelation (DE-AC) approach, we tested three theoretical models with period-doubling bifurcations: Fox, Ricker, and H\'{e}non map. Significantly, Fox model is time-series datasets of the cardiac alternans that exhibit a known period-doubling bifurcation. Additionally, the DE-AC method was tested  on experimental data from chick-heart aggregates that undergo a period-doubling bifurcation. Overall,the DE-AC approach was assessed on both simulated and experimental data, and its performance was compared with the dynamical eigenvalue, lag-1 autocorrelation, and variance.

%
%
%


\section{Methods}

\subsection{Analytical approximation of the autocorrelation function}

As a real eigenvalue is close to $-1$, the system goes through a period-doubling bifurcation. The linearized dynamics that come before each of these bifurcations in their normal form satisfy
\begin{equation}
x_{t+1} = r x_t,
\end{equation}
where, $r$ is the bifurcation parameter and $x_t$ is the displacement from the equilibrium. $\lambda = r$ represents the eigenvalues that determine the stability of the system. Assuming there is tiny additive white noise, the discrete then as follows:
\begin{equation} \label{eq_small_additive_white_noise}
x_{t+1} = \lambda x_t + \sigma \epsilon_t.
\end{equation}
where, $\sigma$ denotes the noise amplitude and $\epsilon_t$ denotes a normal random variable. To calculate the variance of the autoregressive model of order one [AR(1)] process, square both sides of the Eq.~(\ref{eq_small_additive_white_noise}) and take the expectation:
\begin{equation}
\mathbb{E}[x_{t+1}^2] = \mathbb{E}\left[(\lambda x_t + \sigma \epsilon_t)^2\right],
\end{equation}
expanding the right-hand side yields
\begin{equation}
\mathbb{E}[x_{t+1}^2] = \lambda^2 \mathbb{E}[x_t^2] + 2\lambda\sigma \mathbb{E}[x_t \epsilon_t] + \sigma^2 \mathbb{E}[\epsilon_t^2].
\end{equation}
Assuming stationarity, $\mathbb{E}[x_{t+1}^2] = \mathbb{E}[x_t^2]$. Since the normal random variable (white noise term) is independent of $x_t$ with zero mean and unit variance ($\mathbb{E}[\epsilon_t x_t]=0$, $\mathbb{E}[\epsilon^2]=1$), the stationary variance as follows:
\begin{equation}\label{eq.var}
\mathrm{Var}(x) = \mathbb{E}[x_t^2] = \frac{\sigma^2}{1 - \lambda^2}.
\end{equation}
Thus, the variance diverges when the bifurcations approach $\lambda \rightarrow \pm 1$.

Considering the covariance between $x_t$ and $x_{t+1}$ for a stationary process, namely the mean $\mu = 0$, is defined as:
\begin{equation}
\mathrm{Cov}(x_t, x_{t+1}) = \mathbb{E}[(x_t - \mu)(x_{t+1} - \mu)] = \mathbb{E}[x_t x_{t+1}].
\end{equation}
Multiplying Eq.~\ref{eq_small_additive_white_noise} by $x_t$ and taking expectations, gives
\begin{equation}
\mathbb{E}[x_{t+1} x_t] = \lambda \mathbb{E}[x_t^2] + \sigma \mathbb{E}[\epsilon_t x_t].
\end{equation}
Since the noise term $\epsilon_t$ is independent of $x_t$, the second expectation vanishes ($\mathbb{E}[\epsilon_t x_t]=0$), so for the AR(1) process:

\begin{equation}
AC(1) = \frac{\mathrm{Cov}(x_t, x_{t+1})}{\mathrm{Var}(x_t)} = \frac{\mathbb{E}[x_t x_{t+1}]}{\mathbb{E}[x_t^2]} = \lambda.
\end{equation}
Extending this recursively leads to the general expression for the lag-$\tau$ autocorrelation function:
\begin{equation}
ACF(\tau) = \lambda^{|\tau|}.
\end{equation}
Accordingly, the autocorrelation either rises or falls near the period-doubling (flip) bifurcation, depending on the lag time.

\subsection{Theoretical models used for testing}
We employ a model of cardiac alternans that incorporates additive Gaussian white noise, commonly referred to as the Fox model~\cite{bury2023predicting,fox2002period,hall1999prevalence} is described by the following equations
\begin{align}
D_{n+1} &= (1 - \alpha M_{n+1}) \left( A + \frac{B}{1 + e^{-(I_n - C)/D}} \right) + \sigma \epsilon_n, \\
M_{n+1} &= e^{-I_n/\tau} \left[ 1 + (M_n - 1) e^{-D_n/\tau} \right],  \\
I_n &= T - D_n,
\end{align}
where, $D_{n}$ denotes the action potential duration at the $n$-th beat, $M_{n}$ represents a memory variable, and $I_n$ is the rest duration after the action potential. The stimulation period $T$ serves as the control parameter. The parameter $\tau$ denotes the time constant of accumulation and dissipation of memory. The parameter $\alpha$ captures the impact of memory on the action potential duration. Parameters $A$, $B$, $C$ and $D$ influence the shape of the restitution curve. Following Ref.~\cite{fox2002period}, the parameters are fixed as $\tau = 180$, $\alpha = 0.2$, $A = 88$, $B = 122$, $C = 40$, and $D = 28$, which give dynamics consistent with a complex ionic model. The control parameter $T$ decreases linearly over the interval $[300, 150]$, resulting in a period-doubling bifurcation. In this configuration, a period-doubling bifurcation occurs at approximately $T = 200$. $T$ decreases linearly from $300$ to the bifurcation point (approximately $200$) is simulated for $500$ time steps. The noise amplitude is set to $\sigma = 0.05$, and the stochastic term $\epsilon_n$ follows a standard normal distribution with zero mean and unit variance.


To test detection of a period-doubling bifurcation, we use the Ricker model~\cite{bury2020detecting,ricker1954stock,ma2024relaxation,dakos2017elevated} with a harvesting term and additive Gaussian noise. The Ricker model reads as
\begin{align}
x_{t+1} &= x_t e^{r(1-x_t/k)} - F \frac{x_t^2}{x_t^2 + h^2} + \sigma\epsilon_t,
\end{align}
where $x_t$ is the population size at time step $t$, $r$ is the intrinsic growth rate, $k$ is the carrying capacity, $F$ is the harvesting rate, $h$ is a half-saturation constant of the harvesting term, $\sigma$ is the noise amplitude and $\epsilon_{t}$ is a normal random variable with zero mean and unit variance. Parameters are $k = 10$, $F = 0$, $h = 0.75$, $\sigma = 0.1$.
The model exhibits a period-doubling bifurcation at $r = 2.00$, followed by a series of further periodic oscillations to chaos. Forced simulations are run with $r$ increasing linearly on the interval $[0.5, 3.0]$ and null simulations are run with $r = 0.5$.

The H\'{e}non map~\cite{grziwotz2023anticipating,henon1976two} is a discrete-time dynamical system that serves as a simple model of period-doubling bifurcations. This is given by
\begin{align}
x_{t+1} &= 1 - a x_t^2 + y_t + \sigma\epsilon_t,\\
y_{t+1} &= b x_t,
\end{align}
where ($x_t$, $y_t$) represents the state of the system at iteration $t$, $a$ and $b$ are parameters that control the behavior of the system, $\sigma$ is the noise strength and $\epsilon_{t}$ is a standard normal distribution with zero mean and unit variance.  We take $b=0.3$ and $\sigma = 0.1$, which yields a period-doubling bifurcation at $a=0.4$, followed by a series of further periodic oscillations to chaos. Forced simulations are run with $a$ increasing linearly on the interval $[0.1, 0.6]$ and null simulations are run with $a = 0.1$.

\subsection{Experimental data used for testing}


This study employed publicly available datasets derived from experiments involving chick heart cell aggregates undergoing transitions to cardiac arrhythmias~\cite{bury2023predicting,quail2015predicting}. The aggregates were exposed to 0.5~$\mu$mol -- 2.5~$\mu$mol of E4031, a compound known to inhibit the human Ether-$\grave{a}$-go-go-Related Gene (hERG) potassium
channel~\cite{clay1994review}. Following drug administration, some aggregates exhibited alternating in inter-beat intervals, indicative of a period-doubling bifurcation. The inter-beat interval was measured as the time separating consecutive beats and served as the basis for analysis. Comparable transitions are also observed in the human heart, manifesting as T-wave alternans, a phenomenon associated with heightened risk of sudden cardiac death. The onset of the period-doubling bifurcation was identified according to the method of Bury et al.~\cite{bury2023predicting}, where it is marked by the first occurrence of a linear regression slope less than -0.95 over a sliding window of 10 inter-beat intervals on the return map. The dataset comprises 23 time series of aggregates experiencing a period-doubling bifurcation, and 23 null time series of aggregates that did not undergo transformation. Prior to analysis, all time series were preprocessed as described in Bury et al.~\cite{bury2023predicting}, employing a Gaussian filter with a 20-beat bandwidth. Furthermore, early warning indicators were evaluated using a rolling window of 0.5. In this work, we utilized the Python package ewstools~\cite{bury2023ewstools} to smooth and compute generic early warning signals.

\section{Results}

\subsection{Application to  three theoretical models}

\begin{figure*}[h]
\centering
\includegraphics[width=1.0\textwidth]{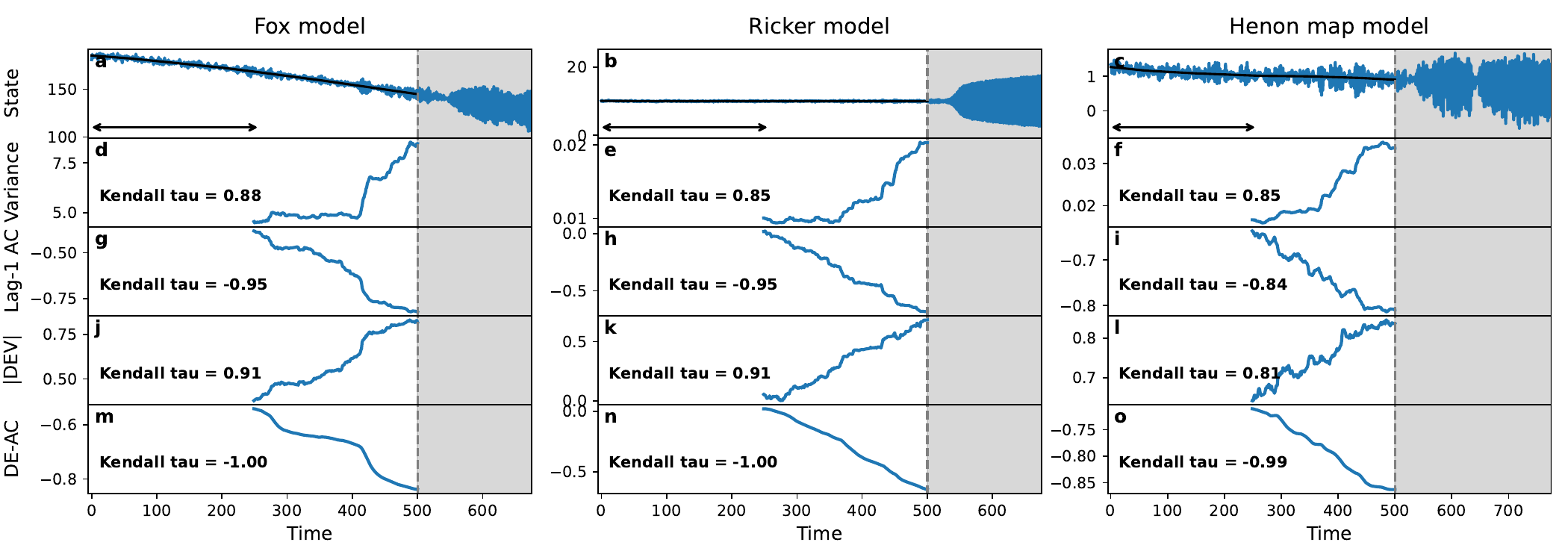}
\caption{\textbf{Trends in indicators preceding the period-doubling bifurcations in three theoretical models.} (\textbf{a}-\textbf{c}) The state trajectory (blue) and its smoothed trace (black) of a simulation undergoing a period-doubling in the Fox, Ricker, and H\'{e}non map  models. The arrow indicates the rolling window (50\% of the time series) applied to calculate EWS. (\textbf{d}-\textbf{f}) Variance (y axis) as a function of simulation time (x axis). (\textbf{g}-\textbf{i}) Lag-1 autocorrelation as a function of simulation time. (\textbf{j}-\textbf{l}) Dynamical eigenvalue ($|$DEV$|$)~\cite{grziwotz2023anticipating} as a function of simulation time. (\textbf{m}-\textbf{o}) Dominant eigenvalue extracted from autocorrelation (DE-AC) as a function of simulation time. Measure the trend of the EWS with Kendall tau. The positive values indicate increasing trends; Negative values indicate decreasing trends. A maximal (minimal) Kendall tau value of $1$ ($-1$) indicates that every subsequent point takes a larger (smaller) value. Kendall tau values show consistent trend in dominant eigenvalue (Kendall tau $ = -1.0$). The vertical dashed line indicates the tipping point (i.e., onset of period-doubling bifurcations), and the gray dashed area delineates the regime following the critical transition.}\label{fig2}
\end{figure*}

We test discrete-time nonlinear dynamical model of cardiac alternans in additive Gaussian white noise, Fox model, representing the alternating cardiac rhythms (i.e., period-doubling bifurcation). In addition, we also validated DE-AC on the Ricker and H\'{e}non map models to test its effectiveness. Figures~\ref{fig2}a--\ref{fig2}c show a period-doubling bifurcation occurring at roughly ``Time = 500", resulting in the beginning of an alternating rhythm. As more of the time series is revealed, we monitor DE-AC, dynamical eigenvalue, lag-1 autocorrelation, and variance. DE-AC, dynamical eigenvalue, lag-1 autocorrelation, and variance are identified as EWS when they exhibit a strong trend. This trend is evaluated quantitatively using the Kendall tau statistic. In three theoretical models, we note an upward trend in variance (see Fig.~\ref{fig2}d--Fig.~\ref{fig2}f), a downward trend in lag-1 autocorrelation (see Fig.~\ref{fig2}g--Fig.~\ref{fig2}i), an increasing trend in $|$DEV$|$ (see Fig.~\ref{fig2}j--Fig.~\ref{fig2}l),  and a decreasing trend in DE-AC (see Fig.~\ref{fig2}m--Fig.~\ref{fig2}o). Our analyses showed that the DE-AC, derived from the analytical form of the lag-$\tau$ autocorrelation function, accurately estimated the dominant eigenvalue of nonlinear dynamical systems and captured its dynamic behavior as it neared a period-doubling bifurcation. As the system approaches a period-doubling bifurcation, the DE-AC value decreases monotonically and tends toward $-1$. This occurs because the real part of the dominant eigenvalue tended to $-1$ (Re($\lambda$) $\to$ $-1$), and its imaginary part approached 0 (Im($\lambda$) $\to$ $0$) in the complex plane (see Fig.~\ref{fig1}a). The dominant eigenvalue ($\lambda$), estimated from the analytical formulation of the lag-$\tau$ autocorrelation function, served as a reliable EWS for predicting the beginning of alternating cardiac rhythms (i.e., period-doubling bifurcation).


\begin{figure}[tbh]
\centering
\includegraphics[width=1.0\textwidth]{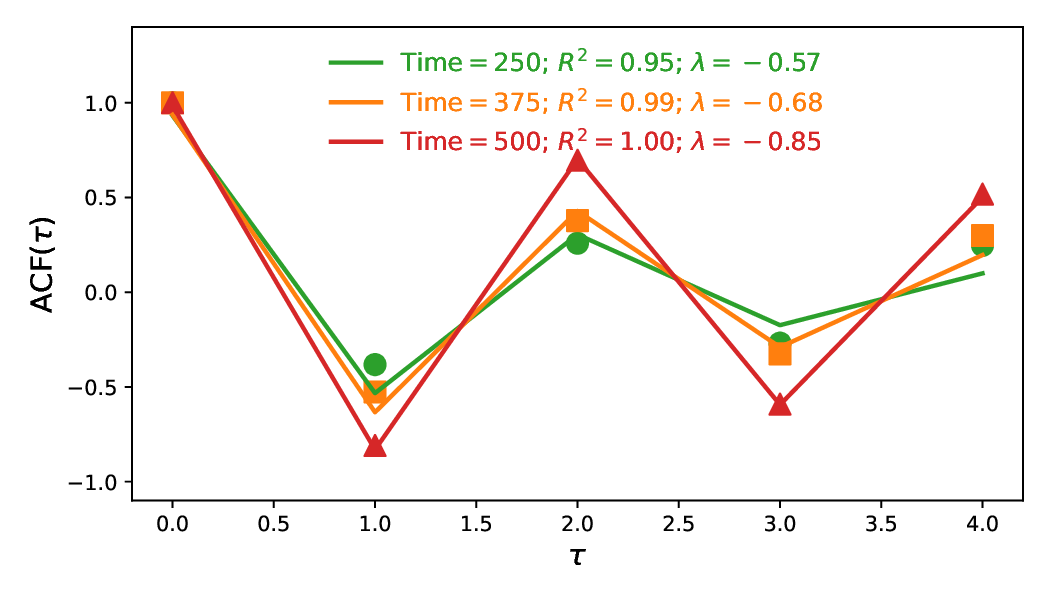}
\caption{\textbf{The lag-$\tau$ autocorrelation function (y axis) for calculated (symbols) and fitted (lines) on the Fox model, as a function of the lag time (x axis).} Each colored line represents a different time and this ``Time" corresponds to those in Fig.~\ref{fig2}. An increasing time indicates approaching the tipping point (i.e., onset of alternating cardiac rhythms). The closer coefficient of determination ($R^2$) is close to $1.0$, which means a higher fitting degree for the indicator. The solid lines match  fitting curves with the form $ACF(\tau)= \lambda^{|\tau|}$, where the dominant eigenvalue $\lambda$ is the free fitting parameters during the fitting process.}\label{fig3}
\end{figure}

Lag-$\tau$ autocorrelation function for three distinct time points (i.e., far (green), medium (orange), and close (red) to the tipping point) before the onset of the alternating cardiac rhythms is shown in Fig.~\ref{fig3}. We observe a strong correlation (coefficient of determination \cite{nagelkerke1991note,chicco2021coefficient}: $R^2 > 0.95$) between the numerical results (symbols) and the fitted results (lines), enabling us to accurately extract the dominant eigenvalue ($\lambda$). It is suggests that the lag-$\tau$ autocorrelation function with period oscillations could be effectively fitted by the following form: $ACF(\tau)=\lambda^{|\tau|}$. In mathematics, the real part of the dominant eigenvalue ($\lambda$) decreases monotonically and tended toward $-1$ (i.e., Re($\lambda$) $\to$ $-1$) when a system approaches period-doubling bifurcation.

\subsection{Applied to chick heart data}


\begin{figure}[tbh]
\centering
\includegraphics[width=1.0\textwidth]{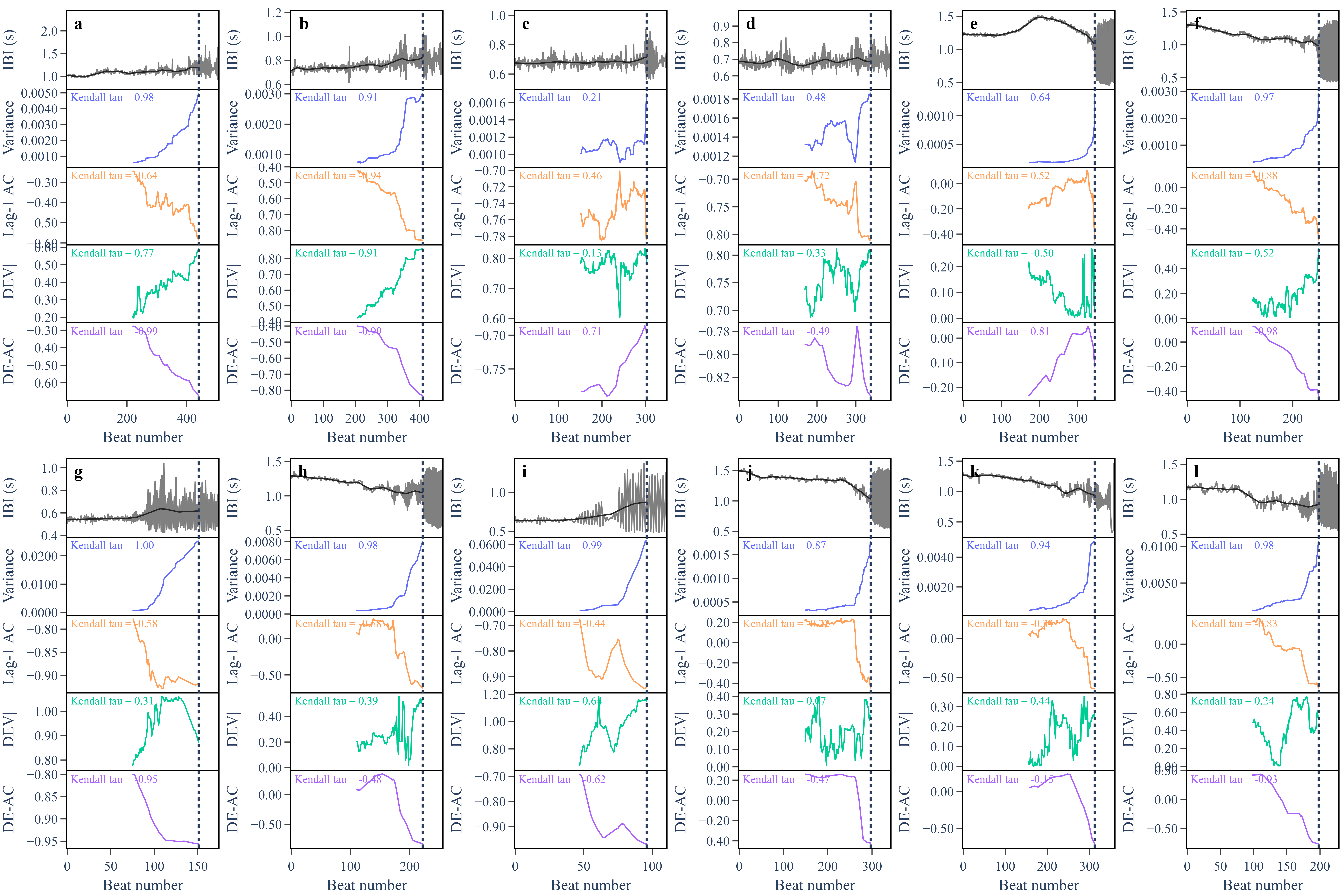}
\caption{\textbf{Trends in indicators preceding alternating cardiac rhythms for chick heart aggregates (IDs 1-12).} (\textbf{a--i}) A period-doubling bifurcation obtained from chick heart data is shown in each panel. (Top) Inter-beat interval (IBI) trajectory (gray) and its smoothed trace (black) from the experimental data. (2nd down, blue) Variance computed using a half-length rolling window from the pre-transition data record. (3rd down, orange) Lag-1 autocorrelation computed using a half-length rolling window from the pre-transition data record. (4rd down, green) Dynamical eigenvalue ($|$DEV$|$) computed using a half-length rolling window from the pre-transition data record. (Bottom, purple) Dominant eigenvalue extracted from autocorrelation (DE-AC) computed using a half-length rolling window from the pre-transition data record. Measure the trend of the EWS with Kendall tau. The beginning of alternating heart rhythms, or onset of period-doubling bifurcation, is marked by the vertical dashed line.}
\label{fig4_1}
\end{figure}

\begin{figure}[tbh]
\centering
\includegraphics[width=1.0\textwidth]{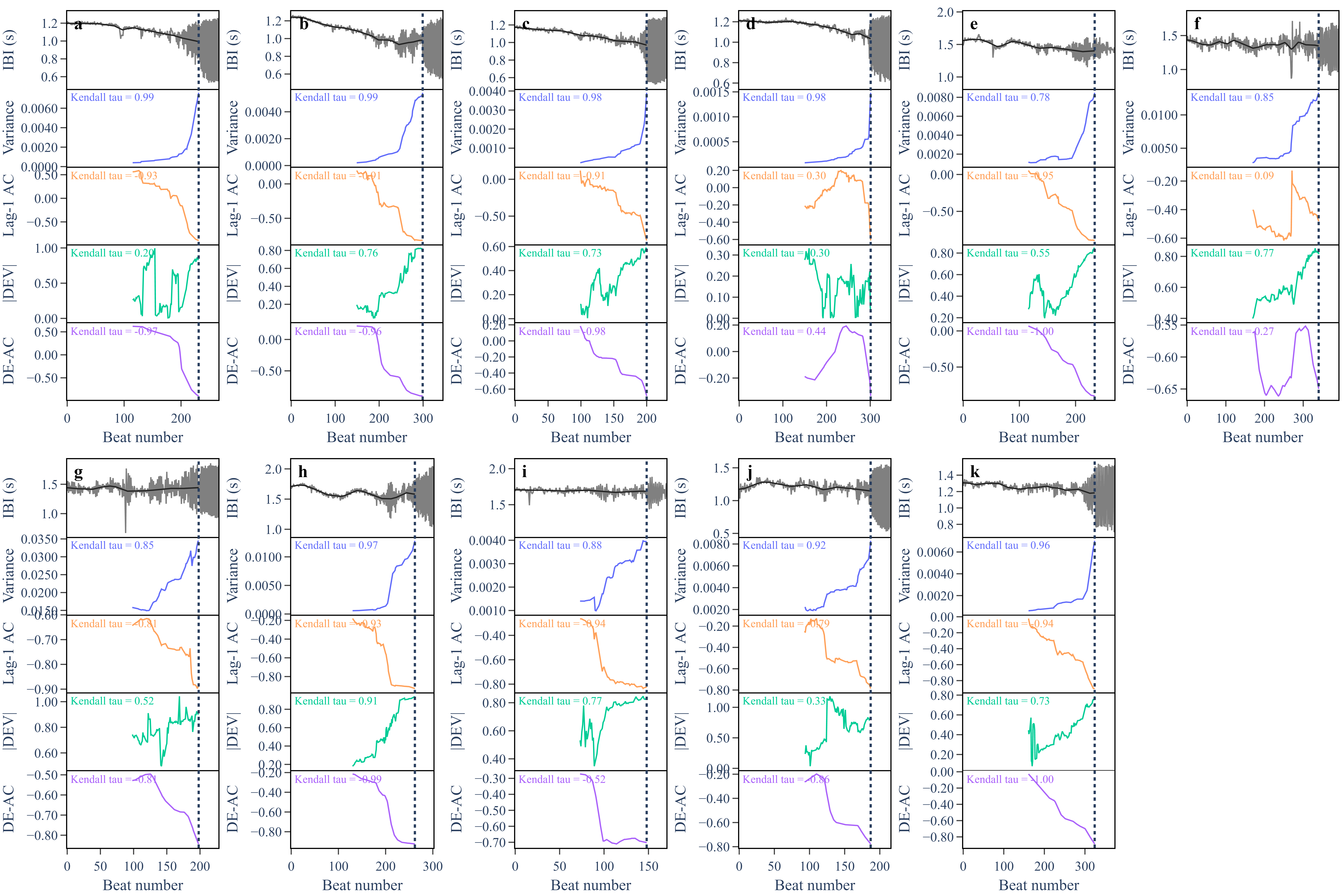}
\caption{\textbf{Trends in indicators preceding alternating cardiac rhythms for chick heart aggregates (IDs 13-23).} (\textbf{a--k}) Each panel shows a period-doubling bifurcation obtained from chick heart data. (Top) Inter-beat interval (IBI) trajectory (gray) and its smoothed trace (black) from the experimental data. (2nd down, blue) Variance computed using a half-length rolling window from the pre-transition data record. (3rd down, orange) Lag-1 autocorrelation computed using a half-length rolling window from the pre-transition data record. (4rd down, green) Dynamical eigenvalue ($|$DEV$|$) computed using a half-length rolling window from the pre-transition data record. (Bottom, purple) Dominant eigenvalue extracted from autocorrelation (DE-AC) computed using a half-length rolling window from the pre-transition data record. Measure the trend of the EWS with Kendall tau. The beginning of alternating heart rhythms, or onset of period-doubling bifurcation, is marked by the vertical dashed line.}
\label{fig4_2}
\end{figure}


\begin{figure}[tbh]
\centering
\includegraphics[width=1.0\textwidth]{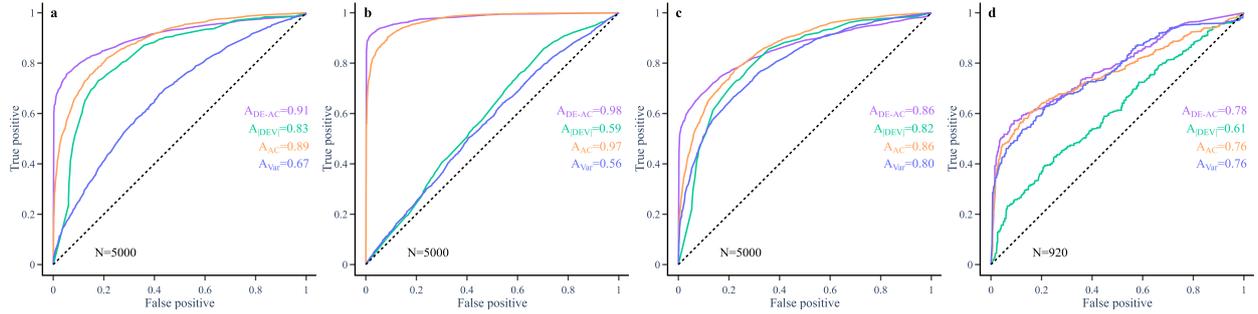}
\caption{\textbf{The receiver operating characteristic (ROC) curves were used to evaluate the performance of an upcoming critical transition for both model and empirical data.} The ROC curves compare the performance of the DE-AC (purple), $|$DEV$|$ (green), variance (blue) and lag-1 autocorrelation (orange). For models (\textbf{a}-\textbf{c}), performance was evaluated using 2500 forced and 2500 null simulations, with varying rates of forcing and noise amplitudes. Predictions employed 70\% of the pre-transition data. For experimental data (\textbf{d}), performance was evaluated over 46 experimental runs, with 20 equal-interval predictions made between 50\% and 70\% of the pre-transition data. Performance is measured by the area under the curve (AUC), which is abbreviated to A. The diagonal dashed line corresponds to a random coin toss baseline with AUC $= 0.5$. (\textbf{a}) Fox model going through a period-doubling bifurcation; (\textbf{b}) Ricker model going through a period-doubling bifurcation; (\textbf{c}) H\'{e}non map model going through a period-doubling bifurcation; (\textbf{d}) Chick heart aggregates going through a period-doubling bifurcation.}
\label{fig6}
\end{figure}

We used embryonic chick heart cell aggregates to predicted the start of alternating cardiac rhythms (see Fig.~\ref{fig4_1} and Fig.~\ref{fig4_2}). In the chick heart data, we notice a downward trend in DE-AC (see Fig.~\ref{fig4_1} and Fig.~\ref{fig4_2}, purple line), an upward trend in $|$DEV$|$ (see Fig.~\ref{fig4_1} and Fig.~\ref{fig4_2}, green line), a decreasing trend in lag-1 autocorrelation (see Fig.~\ref{fig4_1} and Fig.~\ref{fig4_2}, orange line), and an increasing trend in variance (see Fig.~\ref{fig4_1} and Fig.~\ref{fig4_2}, blue line) before the period-doubling bifurcation. Before the stochastic dynamical system undergoes period-doubling bifurcation, increased $|$DEV$|$ and variance, as well as decreased DE-AC and lag-1 autocorrelation, will emerge. This observation is consistent with the theoretical analysis~\cite{bury2023predicting,quail2015predicting,grziwotz2023anticipating,bury2020detecting,wiesenfeld1985noisy}. The dominant eigenvalue ($\lambda$) declined monotonically and tended to $-1$ as the tipping points were approached. The DE-AC (purple line in Fig.~\ref{fig4_1} and Fig.~\ref{fig4_2} ) correctly detects alternating cardiac rhythms (period-doubling  bifurcation) in 18 out of 23 records, namely those that monotonically decrease and tend to $-1$. In alternative scenarios, however, it falsely identifies as an increasing and/or non-monotonic trend. This appears to be associated with an early rise in lag-1 autocorrelation, which may be due to a non-monotonic approach toward the period-doubling bifurcation. In summary, despite noise in real data, the theoretical threshold (Re($\lambda$) $\to$ $-1$) remained meaningful.

\subsection{Performance of early warning signals in model and experimental data}


To assess the performance of the EWS, we evaluate their predictions using both ‘forced’ time series (approaching a bifurcation) and ‘null’ time series (without a bifurcation). For the theoretical models, null time series are generated by fixing the bifurcation parameter. A total of 5000 time series were simulated for the theoretical model. These consisted of five distinct noise intensity values and five different rate of forcing values. Each value was simulated with 100 forced and null time series, respectively. The Kendall tau values serve as discrimination thresholds to create receiver operating characteristic (ROC) curves. For the three theoretical models, we compute the Kendall tau values for DE-AC, dynamical eigenvalue, lag-1 autocorrelation, and variance using 70\% of the pre-transition time series. The DE-AC (see Fig.~\ref{fig6}a--\ref{fig6}c) is a better performer than variance and dynamical eigenvalue, as shown by the AUC score (area under the ROC curve). For the chick heart data, predictions made between 50\% and 70\% of the way through the pre-transition time series yield the highest AUC score for DE-AC (see Fig.~\ref{fig6}d), slightly surpassing variance and lag-1 autocorrelation.

\section{Discussion}

We have derived analytical equations for the lag-$\tau$ autocorrelation function before the period-doubling bifurcation, providing a robust framework to evaluate the dominant eigenvalue of nonlinear dynamical systems. The DE-AC approach distinguishes the onset of period-doubling bifurcations and gives a more sensitive warning for changes in proximity to the tipping point than traditional metrics. We tested the DE-AC approach on simulation data from three theoretical models (Fox, Ricker, and H\'{e}non map) and on experimental data from chick heart aggregates undergoing cardiac arrhythmia. Our results demonstrate that DE-AC correctly recognized the onset of period-doubling bifurcations (alternating cardiac rhythms) and provides early warning signals with higher sensitivity and specificity than three widely used traditional indicators like variance, lag-1 autocorrelation, and dynamical eigenvalue.


The DE-AC offers distinct advantages in terms of computational simplicity and dynamical grounding. Recent advances in large perturbation theory~\cite{lim2011forecasting,ghadami2022data} have demonstrate that monitoring a system's recovery trajectory after a perturbation may yield information about its resilience and distance. Nevertheless, this approach necessitates a controlled environment setting and a system that can safely undergo the large-scale perturbation. Fortunately, data-driven stability monitoring via the Jacobian matrix can identify impending critical transitions using the real parts of eigenvalue~\cite{grziwotz2023anticipating}. Specifically, dynamical eigenvalue method~\cite{grziwotz2023anticipating} is rooted in empirical dynamical modeling (EDM), using state space reconstruction (SSR) to approximate the Jacobian matrix. While versatile, dynamical eigenvalue method requires careful calibration of hyperparameters, such as the embedding dimension ($E$), time lag ($\tau$), window size ($w$), S-map prediction skill ($\rho$), and nonlinear parameter ($\theta$). In contrast, the DE-AC approach does not involve hyperparameters. Because it is derived directly from the relationship between the autocorrelation function and the system's dominant eigenvalue ($ACF(\tau)=\lambda^{|\tau|}$), it bypasses the need for complex SSR. This nature ensures that the extracted eigenvalue is tied to the underlying physical mechanism of critical slowing down rather than being an artifact of the reconstruction process. Consequently, DE-AC is more computationally efficient and more suitable for real-time monitoring in resource-constrained environments, such as wearable medical devices.


Despite its performance, the DE-AC approach has specific limitations. Like conventional early warning signals, the DE-AC can be affected by high noise levels, multi-scale correlations, and sparse sampling~\cite{perretti2012regime}. First, since the derivation is based on the analytical approximations of the autocorrelation function characteristic of period-doubling bifurcation, its direct application to other types of transitions (e.g., Fold or Neimark–Sacker bifurcations) may require further analytical modification. Second, the extraction formula assumes that the system’s fluctuations are governed by linearized dynamics near a steady state. If the system is subject to extremely large stochastic shocks that push it into highly nonlinear regimes far from the attractor, the linear approximation of the dominant eigenvalue may lose precision. Finally, the reliability of the estimate depends on the chosen lag time of the autocorrelation function, which must be long enough to capture the correlation structure but short enough to maintain quasi-stationarity.



Future work should evaluate the DE-AC using high-resolution datasets from natural systems subject to changing environmental conditions over time. Thankfully, improvements in measuring technology are making it easier to obtain high-resolution data in a variety of scientific fields. In ecology, sondes put on lakes to detect chlorophyll content can transmit data every minute, allowing ecologists to predict algae blooms~\cite{pace2017reversal}. In hydrology, open-access global datasets of streamflow and meteorological variables enable daily flood forecasts with 5-day lead times, improving early warning reliability in ungauged and data-scarce watersheds~\cite{nearing2024global}. In medicine, it is now commonplace to monitor patients' physiological data continuously through the use of wearable devices, facilitating the early detection of disease transitions~\cite{tyler2020real}. Therefore, it is very important to have high-resolution data from natural systems.


The DE-AC developed here has important consequences on both science and policy. The DE-AC can infer specific models based on their bifurcation structure, thereby enhancing our understanding of the underlying system. In order to circumvent a regime shift, it is imperative to take action well in advance of the bifurcation~\cite{biggs2009turning}. This means that we necessitates indicators that react swiftly to changes of bifurcation proximity. Given its heightened sensitivity to bifurcation proximity, DE-AC shows promise in enhancing early warning potential. This study contributes to the advancement of techniques for extracting meaningful insights from stochastic dynamics~\cite{boettiger2018noise}, and provides practical tools for immediate application in related analyses.



Previous research on early warning indicators has mainly concentrated on the nonlinear behavior close to fold bifurcation. Although fold bifurcation is important across diverse disciplines, dynamical systems may also undergo critical transitions via other types of bifurcations~\cite{chen2014dynamics}. Therefore, identifying and experimentally validating early warning signals for a broader range of bifurcations is still an open and promising area of research.


\section*{CRediT authorship contribution statement}

Zhiqin Ma: Methodology, Software, Numerical simulations, Formal analysis, Data curation, Visualization, Writing–original draft, Writing–review, Validation, Investigation. Chunhua Zeng: Conceptualization, Methodology, Formal analysis, Supervision, Writing–original draft, Writing–review \& editing, Funding acquisition. Ting Gao: Investigation, Methodology, Supervision, Validation, Writing-review \& editing. Jinqiao Duan: Conceptualization, Validation, Writing-review \& editing.

%
%

\section*{Declaration of competing interest}

The authors declare that they have no known competing financial interests or personal relationships that could have appeared to influence the work reported in this paper.

\section*{Data availability}

Code and instructions to reproduce the analysis are available at the GitHub repository \url{https://github.com/ZhiqinMa/predicting_pd_bifurcation_csf}. Reference of the article containing the data is included in the manuscript and clearly referenced.

\section*{Acknowledgments}

This work was supported by the National Key Research and Development Program of China (Grant No. 2023YFE0109000), the National Natural Science Foundation of China (Grant Nos.~12265017 and 12247205), the Yunnan Fundamental Research Projects (Grant Nos.~202201AV070003 and 202101AS070018), the National Natural Science Foundation of China (Grant Nos.~12401233 and 12141107), the NSFC International Collaboration Fund for Creative Research Teams (Grant No.~W2541005), the Guangdong Provincial Key Laboratory of Mathematical and Neural Dynamical Systems (Grant No.~2024B1212010004), the Yunnan Ten Thousand Talents Plan Young \& Elite Talents Project, Yunnan Province Computational Physics, and Applied Science and Technology Innovation Team.

\bibliographystyle{elsarticle-num}
\bibliography{bibdatabase.bib}



%
%
%

\end{document}